\documentclass[prd,preprintnumbers,nofootinbib,aps,twocolumn]{revtex4-1}
\usepackage{epsfig,amsmath,amssymb}
\usepackage{mathrsfs}
\usepackage[normalem]{ulem}
\usepackage[usenames,dvipsnames]{color}
\usepackage[pagebackref=false, colorlinks=false]{hyperref}

\begin{document}
\title{Can BBH Merger GW Data Constrain Corrections to Bekenstein-Hawking Entropy ? }

\author{Parthasarathi Majumdar}
\email{bhpartha@gmail.com}
\affiliation{School of Physical Sciences, Indian Association for the Cultivation of Science, Kolkata 700032, India.} 

\begin{abstract}

We examine possible additive corrections to the Bekenstein-Hawking (BH) entropy of black holes due to very general classical and quantal modifications of general relativity. In general, black hole entropy is subject to the Generalized Second Law of Thermodynamics. For the case of binary black hole coalescence, the difference in corrections to the inspiral and remnant black hole entropies is shown, within this law, to be bounded by the difference in the corresponding BH entropies. This latter difference has been measured by several groups attempting to validate Hawking's Area Theorem on black hole horizons, by analyzing gravitational wave data from possible binary black hole mergers. The former difference - that between corrections to remnant and inspiral black hole entropies beyond the BH entropy, is thus constrained by a bound measured from observational data. We examine the implications of this constraint for general binary black hole coalesence. If calculated entropy corrections follow the essential pattern of BH entropies dictated by the Hawking Area Theorem, consistency with the observational bound is shown to be guaranteed. If they do not, these corrections are then nontrivially constrained by the observational bound.           
\end{abstract}

\maketitle

\section{Introduction}

The notion that black holes have entropy was proposed in 1973 by Bekenstein \cite{bek73}, inspired by the Laws of Black Hole Mechanics \cite{bch73}. Consistency with Hawking's Area Theorem \cite{haw71} (also called the Second Law of Black Hole Mechanics) requires that such an entropy be a linear function of the cross-sectional area of the event horizon. The justification that such an entropy must exist was enunciated by Bekenstein as the Generalized Second Law (GSL) of thermodynamics, valid for a universe with black holes present. According to this generalization, the total entropy of a black hole plus that of matter outside it can never decrease in any physical process (like some of the matter infalling through the black hole horizon). For vacuum black holes, at the level of classical general relativity, the GSL is identical to the Hawking Area Theorem (HAT) which inspired it. However, generalizations of general relativity as a theory of gravity, both classical and quantal, are likely to lead to corrections to the form of black hole entropy proposed by Bekenstein, metamorphosed into the so-called Bekenstein-Hawking (BH) Area formula, given by $S_{BH} = (A_h/A_{Pl})$ in Boltzmann units ($k_B=1$), where $A_h$ is the cross-sectional area of the black hole horizon, and $A_{Pl} \equiv (G \hbar/c^3)$ is the Planck area. \footnote{The $1/4$ factor traditionally included in the definition of BH entropy will be subsumed here into the definition of  $A_{Pl}$.} These corrections must be restricted by the requirement of the GSL, in any physical process like binary black hole coalescence into a remnant black hole. If such restrictions can be expressed in terms of observationally measureable quantities from Gravitational Wave (GW) data (e.g., in ref.s \cite{LIGO1}-\cite{LIGO7}) arising ostensibly due to  Binary Black Hole coalescence (BBHC), an indirect observational constraint is very likely to emerge in principle, on theoretical modifications to general relativity. 

The aim of this paper is, first to derive a class of inequalities imposed by the GSL on the corrections to BH entropy for different modifications to general relativity, and then to relate these GSL bounds to observational constraints derived from GW data. With regard to the latter, our focus is on results obtained in recent years by several groups \cite{bad06}-\cite{bad22} who have attempted to analyze GW waveform data from a consensus view that these are due to BBH merger observations, made by the LIGO-VIRGO-KAGRA (LVK) consortium. These analyses were driven by the aim to validate the HAT for black hole horizons from GW observations. The affirmative results obtained by these research groups  translate into an observational constraint on the GSL results on the entropy corrections. We think it remarkable that theoretical modifications to general relativity, both classical and quantal, can at all be even partially constrained by currently available GW data, albeit indirectly, assuming the validity of the GSL. 

In an earlier paper \cite{pm24}, corrections to the BH entropy which are logarithmic in the BH entropy itself were considered, in a specific background-independent theory of quantum gravity known as Loop Quantum Gravity (LQG). Within the broad tenets of LQG, these logarithmic corrections are seen to be fully consistent with the results of the above analyses of GW data which affirm the HAT, to the level of accuracy that such data analyses entail. The demonstration of this consistency does not, by any means, require that the data be accurate at the Planck length level at which the logarithmic entropy corrections are often construed to be physically relevant. The reason for this rather extraordinary outcome of our earlier work is to be found in the algebraic sign of the entropy corrections in LQG, rather than the magnitude of these corrections. It was pointed out in ref. \cite{pm24} that logarithmic entropy corrections with the {\it opposite} algebraic sign will be nontrivially restricted by the observational bound. Thus, GW data analyses results indirectly have the power to distinguish quantum gravity approaches leading to logarithmic entropy corrections of different algebraic signs. 

The present letter presents a far broader generalization of this genre of arguments, not restricting it to any specific quantal modification of classical GR. The formal GSL-based inequality, in terms of the observational constraint, is shown to be valid for {\it any additive} correction to BH entropy, irrespective of the origin of such correction. If the entropy correction for any black hole is a function of not just the horizon area (BH entropy) of that black hole, but also other parameters(`charges ${\bf q}= \{q_I|I=1,2, \dots \}$), then for astrophysical black holes (with macroscopic horizons ($S_{BH} >> 1$)) it can be expanded in inverse powers of the BH entropy, with the expansion coefficients being functions of  the charges. Under some very natural assumptions on the nature of these coefficients, we derive the precise conditions for which consistency of the GSL-based inequality with the HAT measurements is guaranteed. Violation of our derived conditions would then imply non-trivial constraints on the entropy corrections from the HAT analyses of observational data. Thus, observational data from BBHCs can indeed constrain, in some cases, theoretically calculated corrections to the BH entropy of black holes.       
        
\section{Generalized Second Law}

Appealing to ref. \cite{lieb99}, the idea of the horizon area $A_h$ as an `energy coordinate', of which the black hole entropy is a function ($S_{bh}=S_{BH}(A_h)$) and the idea of  `adiabatic accessibility' can be invoked to argue that if $\Sigma_1$ is a spatial hypersurface chosen to lie in the future of spatial hypersurface $\Sigma_2$, and the corresponding cross-sectional areas of the Isolated Horizons are $A_{h1}$ and $A_{h2}$ (with $A_{h1} > A_{h2}$ by the HAT) then one should expect $S(A_{h1}) > S(A_{h2})$. If there is matter outside the black hole which can be accreted into it, then one would infer the Generalized Second Law (GSL) of Thermodynamics 
\begin{eqnarray}
S_{bh1} + S_{out1} \geq S_{bh2} + S_{out2} \label{gsl} 
\end{eqnarray}   
Thus, in a universe with black holes, the total entropy comprising the gravitational entropy and the matter entropy outside the horizon can never decrease in any physical process. 

Now imagine a BBHC of an inspiralling pair of black holes into a remnant black hole with emission of gravitational waves; the GSL now assumes the form
\begin{eqnarray}
S_{bhr} + S_{GW} \geq S_{bh1} + S_{bh2} \label{bbhg}
\end{eqnarray}
The subscript $r$ stands for `remnant'. The entropy associated with radiant GW from a generic BBHC has been estimated earlier \cite{pmar21}; it has been shown that $S_{GW} << S_{bhr}, S_{bh1}, S_{bh2}$, and hence can be dropped from further considerations. In this form, it is clear that for general relativity alone, semiclassical considerations imply that $S_{bh} = S_{BH}$;  in this case the GSL has an identical content to that of the HAT. However, if modifications to general relativity, either due to classical considerations like  theory of gravity with higher order derivatives of the metric, or due to various possible {\it quantal} modifications, are included, corrections to $S_{BH}$ would in general arise. The GSL would then have nontrival implications on the importance of these corrections for a BBHC astrophysical event.  

\section{Consequences for Additive Entropy Corrections}

We restrict our considerations to modifications of general relativity such that the entropy corrections to $S_{BH}$ arise additively 
\begin{eqnarray}
S_{bh} = S_{BH} + s_{bh} \label{add}
\end{eqnarray} 
Then, discarding $S_{GW}$ from the inequality (\ref{bbhg}), we obtain, for BBHC, the inequality 
\begin{eqnarray}
\Delta S_{BH} & > & -\delta s_{bh} \label{ineq} \\
\Delta S_{BH} & \equiv & S_{BHr} - (S_{BH1} + S_{BH2}) \label{BHdef} \\
\delta s_{bh} & \equiv & s_{bhr} - (s_{bh1} + s_{bh2})~. \label{corr}
\end{eqnarray} 

The HAT implies categorically that $\Delta S_{BH} > 0$, so the inequality (\ref{ineq}) now has the constraint that the combination $\delta s_{bh}$ of the corrections is accordingly constrained
\begin{eqnarray}
-\delta s_{bh} < \Delta S_{BH} \label{ineq2}
\end{eqnarray} 
We now make the following important observation : if it so turns out that the calculated correction $\delta s_{bh} > 0$, then the constraint implied in (\ref{ineq2}) is automatically satisfied, so long as the HAT is valid. On the other hand, the same validity of the HAT imposes a nontrivial constraint on $s_{bh}$ if it turns out that $\delta s_{bh} < 0$. Thus, the validity of the HAT clearly distinguishes between two classes of theoretically calculated corrections to the $S_{BH}$ in the case of a BBHC. If the corrections to $S_{BH}$ follow a pattern dictated by the HAT, the constraint is a tautology. If not, we have a non-trivial constraint on the size of the corrections.This is rather remarkable given that the modifications of classical general relativity, leading to the corrections, go beyond the tenets of general relativity. We need to recall of course that this entire result hinges on the validity of the GSL; although in our earlier paper \cite{pm24} we have only provided a heuristic justification as to why the GSL ought to hold. Intuitively, however, the fact that the Second Law of Thermodynamics has no recorded violations, lead us to trust that the GSL is valid for all situations which start from and resolve into equilibrium conditions such as a stationary black hole. 

Fortunately for us, the validity of the HAT is now established from observed data, within certain levels of accuracy. Several research groups \cite{bad06}-\cite{bad22} have performed detailed analyses of the the waveforms of gravitational waves from BBHCs observed by the LVK consortium in the two observations GW150914 and GW170714. In these papers, the inspiral and ringdown waveforms of the LVK data for the mergers mentioned, have been analysed separately in the time domain, so as not to be cluttered with possible interference between the two phenomena. Waveforms generated by numerical relativity analyses of the entire Inspiral-Merger-Ringdown (IMR) phases are also considered for comparison. Subtleties associated with the choice of time intervals for the end of the inspiral phase and the start of the ringdown phase, i.e., the time truncations involved, have been resolved by some of the research groups \cite{teu19a} - \cite{isi21}, where the role of inclusion of overtones in the ringdown waveforms have been meticulously analysed. In these subclass of analyses, validity of the No-Hair Theorem from LVK data is an essential feature of the background leading to the HAT validation. However, for the purpose of this letter, it suffices to consider the validity of the HAT in our terms, namely $(\Delta S_{BH})_{obs} > 0$, which can be taken as a consensus result of the HAT analyses of GW data.      
 
This implies that for additive calculated entropy corrections $\delta s_{bh} > 0$, consistency with observational data is guaranteed, while if the calculated corrections have the opposite sign : $\delta s_{bh} < 0$, then the HAT analyses of observational data does impose a non-rivial constraint on the magnitude of these corrections. This binary may be held in view whenever classical or quantal modifications of GR are being considered.  

\section{Further implications on nature of entropy corrections}
\subsection{Logarithmic Corrections}

For astrophysical relevance, we restrict our attention in this letter to black holes with {\it macroscopic} horizon areas, i.e., $S_{BH} >> 1$. The entropy corrections for such black holes 
\begin{eqnarray}
s_{bh} = s_{bh}(S_{BH}, {\bf q}) \label{entc}
\end{eqnarray}
where ${\bf q} = \{ q_I|I=1,2, \dots \} $ characterize black hole parameters (`charges') beyond the GR parameters of mass and spin which characterize $S_{BH}$. For astrophysical black holes $s_{bh}$ admits the expansion
\begin{eqnarray}
s_{bh} = s_0 ({\bf q}) \log S_{BH} + \sum_{n=1}^{\infty} s_n({\bf q}) S_{BH}^{-n} \label{exp}
\end{eqnarray}  
For $S_{BH} >>> 1$, we can ignore the inverse power law corrections compared to the leading logarithmic correction, with the caveat that the coefficients $s_0, s_n$ are bounded for all charges. Given that (\ref{exp}) and the subsequent discarding of power law corrections should hold for the inspiralling pair as well as the remnant black holes, it is easy to see that eqn. (\ref{ineq2}) leads to the inequality
\begin{eqnarray}
\log \left[\frac{S_{BH1}^{s_{01}} S_{BH2}^{s_{02}}}{S_{BHr}^{s_{0r}}} \right] < 0 < (\Delta S_{BH})_{obs} \label{logq}
\end{eqnarray}
where the inequality on the {\it rhs} is the observational evidence from the HAT analyses of data. The coefficients $s_{01,2} = s_{01,2}({\bf q}_{1,2}), s_{0r} = s_{0r}({\bf q}_r)$, and can be expressed as $s_{01,2} = |s_{01,2}| sig(s_{01,2}), s_{0r} = |s_{0r}| sig(s_{0r})$.
 We now make the assumptions 
\begin{enumerate}
\item The algebraic sign $sig(s_{01,2,r})$ is the same for {\it all} black holes, inspiral and remnant. 
\item The magnitudes of the expansion coefficients are roughly of the same order.
\item $S_{BH1} \cdot S_{BH2} > S_{BHr}$, as is true for GW150914. 
\end{enumerate}
Substitution into eqn (\ref{logq}) leads to 
\begin{eqnarray}
sig (s_0) \log \left[\frac{S_{BH1}^{|s_{01}|} S_{BH2}^{|s_{02}|}}{S_{BHr}^{|s_{0r}|}} \right] < 0. \label{logq-}
\end{eqnarray}
Eqn (\ref{logq-}) immediately requires that $sig(s_0) = -$, i.e, if the log-corrected black hole entropy is {\it less than} the BH entropy for all black holes, the consistency of the entropy corrections in a BBHC event with observational data is {\it guaranteed}, without any constraints. 

On the other hand, if $sig (s_0) = +$, together with assumptions 1-3, we simply restate the constraint (\ref{logq}) 
\begin{eqnarray}
\log \left[\frac{S_{BH1}^{s_{01}} S_{BH2}^{s_{02}}}{S_{BHr}^{s_{0r}}} \right] < (\Delta S_{BH})_{obs} \label{logq+}
\end{eqnarray}
i.e., in this case, the log-corrected entropy exceeds the BH entropy and as such, the corrections for a BBHC event are bounded from above by the observational result stemming from the HAT analyses of GW data. Since logarithmic corrections to the BH entropy typically arise in quantum gravitational formulations of black hole entropy, it is indeed remarkable that various quantum gravity approaches to calculate these corrections bifurcate into two classes that are either consistent with observations without constraint on the magnitudes, or have constraints on the magnitudes in terms of the  observational result. Given that the observational result used here does not involve any actual formulation where log-corrections to the BH entropy of black holes are computed, the conclusions in this paragraph are, in a sense, {\it unexpected}.     
 
\subsection{Power law corrections}

For classical modifications of GR, as in the addition  of terms of higher powers of curvature to the Einstein-Hilbert-Lorentz action including the Lovelock and the Horndeski theories of gravity, modifications to the BH entropy, as in the Wald entropy formulation \cite{wald92}, is unlikely to involve logarithmic corrections to the entropy. For this purpose, we can set the log-correction coefficient $s_0 = 0$ for all black holes in a BBHC event. Restricting to the leading correction to the BH entropy involving $S_{BH}^{-1}$, and demanding unconstrained consistency with (\ref{ineq2}), we obtain the inequality
\begin{eqnarray}
s_{11} S_{BH1}^{-1} + s_{12} S_{BH2}^{-1} < s_{1r} S_{BHr}^{-1} .\label{pow}
\end{eqnarray}    
From the HAT $\Delta S_{BH} > 0$, we obtain $S_{BHr}^{-1} < (S_{BH1} + S_{BH2})^{-1}$, which leads to the inequality 
\begin{eqnarray}
s_{1r} > s_{11} + s_{12}. \label{pow1}
\end{eqnarray}
If this inequality is satisfied for any classical modification of GR, consistency with the observational result discussed here is guaranteed, without any further restrictions on the expansion coefficients $s_1$ .  We have not made any assumptions at all on the sort of parameters or charges that may actually characterize the black hole solution in the proposed classical modification of GR.  

On the other hand, if the inequality (\ref{pow1}) is not satisfied in the modified formulation of a gravity theory beyond GR, i.e., $s_{1r} < s_{11} + s_{12}$, then the power law corrections are restricted in magnitude by the inequality
\begin{eqnarray}
\frac{s_{11}}{S_{BH1}} + \frac{s_{12}}{S_{BH2}} - \frac{s_{1r}}{S_{BHr}} < (\Delta S_{BH})_{obs} .\label{pow2}
\end{eqnarray} 

\section{Discussion}

Our {\it model-independent} analysis above did not use any specific modification of GR, either classical or quantal. Perhaps therein lies the strength of the ensuing conclusions presented here. In the case of unconstrained consistency of the entropy corrections with the observational result, it is not exactly clear if there is a general reason for this. Of course, for the logrithmic corrections, unconstrained consistency with observation happens if the corrected entropy is smaller that the BH entropy. Since black hole entropy counts the gravitational microstates, this would imply that a smaller number yields the consistency. Such a lowering of the degrees of freedom would emerge if the counting leading to the BH entropy overcounts microstates related by a symmetry. E.g., in the case of logarithmic corrections arising from Loop Quantum Gravity \cite{pm24}, the imposition of the Gauss law constraint corresponding to local rotational invariance of physical states, is known to lead to precisely the condition for consistency derived model-independently in this letter. This is not however to claim that imposing local rotational invariance always leads to the desired result, although there is positive likelihood of that being the origin of the consistency. This needs a deeper investigation. Similarly, whether at the classical level the consistency of the power law corrections with observation is guaranteed by the inequality (\ref{pow1}) because this inequality results from the underlying local rotational invariance, is not clear at this point.

On the other hand, how stringent are the constraints in case that consistency with observation is not guaranteed ? Can they be improved with better statistics and future runs of Advancd LIGO ? Perhaps, but at this stage of the science, it is doubtful if more can be said about this. Our model-independent approach is perhaps not quite adequate yet to rule out any particular proposal for modification of GR, be it classical or quantal. But our results here may point out prospective areas of tension of proposed modiciations of GR with observation. This, given that this area is under a lot of active investigation, is not without significance for the future.  

\section{Acknowledgements}

I thank Will M. Farr, Saul A. Teukolsky, Lieke van Son and especially Soumendra K. Roy for many illuminating discussions. This work evolved to its present form because of the hospitality that the author received at the Center for Computational Astrophysics at the Flatiron Institute, New York, USA and the Center for Space Physics at Cornell University, Ithaca, New York, USA. I am most grateful to Professors Farr and Teukolsky for their generous hospitality. The idea for this work emerged during a presentation at the Leonard E. Parker Center for Gravitation, Cosmology and Astrophysics, University of Wisconsin-Milwaukee, last Spring. I thank Jolien Creighton, John Friedman and especially Anarya Ray for many wonderful discussions, and Professor Creighton and the CGCA for most gracious hospitality.

\end{document}